\documentclass[twocolumn,epjc3]{svjour3}  
\smartqed  % flush right qed marks, e.g. at end of proof
\RequirePackage{graphicx}
\journalname{Eur. Phys. J. C}
\def\beq{\begin{equation}}
\def\eeq{\end{equation}}

\begin{document}

\title{Particle scattering by a test fluid on a Schwarzschild spacetime: the equation of state matters%\thanksref{t1}
}

\author{Donato Bini\thanksref{e1,addr1,addr2}
\and
        Andrea Geralico\thanksref{e2,addr2,addr3} 
\and
Sauro Succi\thanksref{e3,addr1}
}

\thankstext{e1}{e-mail: binid@icra.it}
\thankstext{e2}{e-mail: geralico@icra.it}
\thankstext{e3}{e-mail: s.succi@iac.cnr.it}


\institute{Istituto per le Applicazioni del Calcolo \lq\lq M. Picone," CNR, I-00185 Rome, Italy\label{addr1}
           \and
           ICRA, University of Rome \lq\lq La Sapienza," I-00185 Rome, Italy\label{addr2}
           \and
           Physics Department, University of Rome \lq\lq La Sapienza," I-00185 Rome, Italy\label{addr3}
%           \emph{Present Address:} if needed\label{addr3}
}

\date{Received: date / Accepted: date}

\maketitle

\begin{abstract}
The motion of a massive test particle in a Schwarzschild spacetime surrounded by a  
perfect fluid with equation of state  $p_0= w \rho_0$ is investigated. 
Deviations from geodesic motion are analyzed as a function of the parameter 
$w$, ranging from $w=1$ which corresponds to the case of massive free scalar fields, down into 
the so-called ``phantom" energy, with $w<-1$.
It is found that the interaction with the fluid distribution leads to capture (escape) of the particle trajectory in the case $1+w>0$ ($<0$), respectively. 
Based on this result, it is argued that inspection of the trajectories of test particles in the vicinity
of a Schwarzschild black hole may offer a new means of gaining insights into the nature of cosmic matter.

\keywords{Scattering of particles \and Schwarzschild spacetime \and Poynting-Robertson-like effects}  %\and Second keyword \and More}
\PACS{04.20.Cv } %\and PACS code2 \and more}
\end{abstract}

\section{Introduction}
\label{intro}

Material bodies moving within a surrounding fluid are typically subject to 
a force associated with the fluid-body interaction. 
For instance, in the case of a photon gas this force acts like a friction contrasting the particle 
motion, as first considered by Poynting \cite{poynting} and Robertson \cite{robertson}.
The force term entering the equations of motion was given there by the 4-momentum density of radiation observed in the particle's rest frame with a multiplicative constant factor expressing the strength of the interaction itself.
In the present paper we follow the same approach to investigate the general features of particle motion in a Schwarzschild spacetime region filled with a fluid with equation of state $p_0=w \rho_0$.
The parameter $w$ characterizes different kind of fluids, from massive free scalar fields ($w=1$), to 
non-relativistic dust ($w=0$), all the way down deep into the so-called \lq\lq phantom" energy ($w<-1$).
As it is well known from mathematical cosmology \cite{islam}, the parameter $w$ has a 
profound influence on the time evolution of the Universe. For instance, in the Friedmann-Robertson-Walker (FRW)
models it determines the power law exponent of the scale factor as a function of time.  
According to our analysis, which relies on an explicit form of the 
fluid-particle interaction, the parameter $w$ plays a major role also in discriminating 
among different types (escape versus capture) of particle motion, for a given black hole spacetime metric. 
In particular, it is shown that the condition $w+1=0$ (the case of a cosmological constant) identifies
with geodesic motion in the Schwarzschild metric, and sets 
the borderline between capture (positive sign) and escape (negative sign), respectively.
This sharp borderline could in principle lead to measurable effects, hence
new hints on the nature of the cosmological matter, based on the observation
of the motion of test particles embedded in the cosmological fluid surrounding
a Schwarzschild black hole.

\section{Test fluid superposed to a Schwarzschild spacetime}

We begin by writing the Schwarzschild metric in the standard form
\beq
\label{sch}
ds^2=-N^2dt^2+N^{-2}dr^2+r^2(d\theta^2+\sin^2\theta d\phi^2)\,,
\eeq
with the \lq\lq lapse function" $N$ given by
\beq
N=\sqrt{1-\frac{2M}{r}}\,,
\eeq
where $M$ is the mass of the black hole. 
The metric (\ref{sch}) is static, i.e., $\xi^\mu=(\partial_t)^\mu$ is a hypersurface-forming timelike Killing vector. 
Observers at rest with respect to the coordinates (or \lq\lq static observers") have their $4$-velocity vector 
aligned along the Killing direction itself, namely
\beq
u  =N^{-1}\partial_t\,.
\eeq
An orthonormal frame $e_{\hat a}$  ($a=1,2,3$) adapted to them is given by 
\begin{eqnarray}
\label{frame}
e_{\hat r}=N\partial_r\,,\quad e_{\hat \theta}=\frac{1}{r}\partial_\theta\,,\quad e_{\hat\phi}=\frac{1}{r\sin\theta}\partial_\phi\,.
\end{eqnarray}

For later use, let us introduce co-rotating $(+)$ and counter-rotating $(-)$ circular timelike geodesic orbits 
on the equatorial plane $\theta=\pi/2$, characterized by 
the \lq\lq Keplerian" $4$-velocity
\beq
U_K=\gamma_K (u\pm \nu_K e_{\hat \phi})=\Gamma_K (\partial_t \pm \zeta_K \partial_\phi)\,,
\eeq
where $\gamma_K=(1-\nu^2_K)^{-1/2}$ and $\Gamma_K=\gamma_K/N$, with
\beq
\label{kepler}
\nu_K=\sqrt{\frac{M}{r-2M}}\,,\qquad \gamma_K\,=\sqrt{\frac{r-2M}{r-3M}}\,.
\eeq
These orbits only exist  for $r>3M$.
Next, let us consider a (test) perfect fluid, superposed to the Schwarzschild spacetime, described 
by the following stress-energy tensor \cite{mtw}
\beq
\label{tensorefotoni2}
T^{\mu \nu}(r)= \rho_0(r)\Big[(1+w)u^\mu u^\nu +w  g^{\mu \nu} \Big]\,,
\eeq
i.e., with equation of state $p_0(r)=w \rho_0(r)$, where $w$ is a constant characterizing 
the type of cosmic fluid, as detailed below. 
The form of $\rho_0(r)$ is obtained by integrating the equations $\nabla_\nu T^{\mu\nu}=0$; in this 
case, due to spherical symmetry of the background, the solution of these equations is easily obtained, i.e.,
\beq
\rho_0(r)=  \rho_0^\infty \, N^{-\frac{(w+1)}{w}}\,,
\eeq
where $\rho_0^\infty$ is a constant representing the value of the energy density at infinity.
Noted values of $w$ in the literature (see, e.g., Refs. \cite{sahni,vikman,baum}) are
as follows:
\begin{itemize}
  \item[] $w=1$: massive free scalar field;
  \item[] $w=1/3$: radiation;
  \item[] $w=0$: dust (non-relativistic matter);
  \item[] $-1/3 < w < -1$: quintessence;
  \item[] $w=-1$: cosmological constant;
  \item[] $w<-1$: phantom energy. 
\end{itemize}

As is well known, each of these values corresponds to a different growth of the cosmological
scale factor in the simple FRW model, that is
$R(t)=R(t_0) (t/t_0)^{\frac{2/3}{(1+w)}}$. This shows that the capture scenario, i.e.,
$1+w>0$, corresponds to initial time singularities (Big-Bang), whereas the escape scenario, i.e.,
$1+w<0$, associates with finite-time future singularities (Big-Rip) (see Ref. \cite{Brevik} and references therein). 

\section{Scattering of particles}

Based on the energy momentum tensor (\ref{tensorefotoni2}), we model the force acting 
on a massive particle with $4$-velocity $U^\alpha=dx^\alpha/d\tau$, in the form of a 
linear drag term, proportional to the particle velocity 
\beq
\label{frad_gen}
f_{\rm (scat)}(U)_\alpha=- \sigma P(U)_{\alpha \beta} T^\beta{}_\mu U^\mu\,,
\eeq 
where 
\beq
\label{Udef}
U=\gamma (u+\nu^{\hat a}\, e_{\hat a})\,,\quad \gamma=(1-\delta_{\hat a\hat b}\nu^{\hat a}\nu^{\hat b})^{-1/2}\,,
\eeq
$P(U)=g+U\otimes U$ projects orthogonally to $U$ and $ \sigma$ is a numerical coefficient. 
This expression for the scattering force has been already used in the literature to study radiation 
scattering \cite{poynting,robertson,poy1,poy2,vaidyaPR,bg_scatt}. 
In that case, $\sigma$ models the absorption  and consequent re-emission of radiation by the particle. 
Here, we extend its form in such a way as to cover more general situations.
The entire discussion about deviations from geodesic motion given below relies on this choice of the force.
Although by no means unique, the definition (\ref{frad_gen}) has the merit of  
mathematical simplicity, as combined with a clear physical basis.

The motion of a massive particle is thus governed by the equations
\beq
\label{eq_mot_gen}
m a(U)=f_{\rm (scat)}(U)\,,
\eeq
where $a(U)=\nabla_U U$ denotes the particle $4$-acceleration with frame components 
\begin{eqnarray}
\label{fouracc}
a(U)^{\hat t} 
&=& \frac{d\gamma}{d\tau}+\frac{\gamma^2N}{r}\nu_K^2\nu^{\hat r}\,, \\
a(U)^{\hat r}
&=& \frac{d (\gamma \nu^{\hat r})}{d\tau}-\frac{\gamma^2 N}{r}(\nu^{\hat \theta}{}^2+\nu^{\hat \phi}{}^2-\nu_K^2)\,,\nonumber \\
a(U)^{\hat \theta}&=& \frac{d (\gamma \nu^{\hat \theta})}{d\tau}+\frac{\gamma^2}{r\sin\theta}[N\sin \theta \nu^{\hat r}\nu^{\hat \theta}-\cos \theta \nu^{\hat \phi}{}^2]\,,\nonumber \\
a(U)^{\hat \phi}&=& \frac{d (\gamma \nu^{\hat \phi})}{d\tau}+\frac{\gamma^2}{r\sin\theta}[N\sin \theta \nu^{\hat r}+\cos \theta \nu^{\hat \theta}]\nu^{\hat \phi}\,.\nonumber
\end{eqnarray}
The quantities $\nu_K$ and $\gamma_K$ have been introduced in Eq. (\ref{kepler}).
From the orthogonality condition $U\cdot a(U)=0$, the following relation holds
\beq
\label{formula_acc}
a(U)^{\hat t}=\nu_{\hat r} a(U)^{\hat r}+\nu_{\hat \theta} a(U)^{\hat \theta}+\nu_{\hat \phi} a(U)^{\hat \phi}\,.
\eeq
In the sequel, we shall focus on the special case of a particle confined to the equatorial 
plane ($\theta=\pi/2$), namely with $\nu^{\hat \theta}=0$.
In this case, $a(U)^{\hat \theta}=0$ and the remaining components simplify to
\begin{eqnarray}
\label{fouracceq}
a(U)^{\hat t}
&=& \frac{d\gamma}{d\tau}+\frac{\gamma^2N}{r}\nu_K^2\nu^{\hat r}\,,\nonumber \\
a(U)^{\hat r}
&=& \frac{d (\gamma \nu^{\hat r})}{d\tau}-\frac{\gamma^2 N}{r}(\nu^{\hat \phi}{}^2-\nu_K^2)\,,\nonumber \\
a(U)^{\hat \phi}&=& \frac{d (\gamma \nu^{\hat \phi})}{d\tau}+\frac{\gamma^2 N }{r} \nu^{\hat r}\nu^{\hat \phi}\,.
\end{eqnarray}
The general equatorial motion is thus fully described by the following equations
\begin{eqnarray}
\label{eq_nus}
\frac{d \nu^{\hat r}}{d\tau}
&=& \frac{1}{\gamma}\left(\frac{a(U)_{\hat r}}{\gamma_r^2}- a(U)_{\hat \phi}\nu^{\hat r}\nu^{\hat \phi}\right)\nonumber\\
&&+
\frac{\gamma N}{r}[\nu^{\hat \phi}{}^2 -\nu_K^2 (1-\nu^{\hat r}{}^2)]\,,\nonumber \\
\frac{d  \nu^{\hat \phi}}{d\tau}
&=& 
\frac{1}{\gamma}\left(\frac{a(U)_{\hat \phi}}{\gamma_\phi^2}- a(U)_{\hat r}\nu^{\hat r}\nu^{\hat \phi}\right)
-\frac{\gamma N}{\gamma_K^2 r}\nu^{\hat r}\nu^{\hat \phi}\,,
\end{eqnarray}
where $\gamma_r=1/\sqrt{1-\nu^{\hat r}{}^2}$ and $\gamma_\phi=1/\sqrt{1-\nu^{\hat \phi}{}^2}$, once the acceleration components are replaced by the force components according to
\begin{eqnarray}
a(U)^{\hat r}&=& \frac{1}{m}f_{\rm (scat)}(U)^{\hat r}=-\frac{\sigma}{m} (w+1) \rho_0^\infty\gamma^3 \nu^{\hat r} \,, \nonumber \\
a(U)^{\hat \phi}&=& \frac{1}{m}f_{\rm (scat)}(U)^{\hat \phi}=-\frac{\sigma}{m} (w+1) \rho_0^\infty\gamma^3 \nu^{\hat \phi}  \,.
\end{eqnarray}
The evolution equations for the radial and azimuthal coordinates follow directly from the definition of $U$ (see Eq. (\ref{Udef})), i.e.,
\begin{eqnarray}
\frac{d r}{d\tau}&=& \gamma N \nu^{\hat r}\,,\qquad
\frac{d \phi}{d \tau}=\frac{\gamma}{r} \nu^{\hat \phi}\,.
\end{eqnarray}

We note that when $w=-1$, the force term $f_{\rm (scat)}(U)$ vanishes identically, implying geodesic motion.
Therefore, if the superposed fluid corresponds to a cosmological constant term in the Einstein's equations, the 
interaction between particles and gas cannot be modeled by Eq. (\ref{frad_gen}). 
This is exactly the case of the de Sitter spacetime.
In fact, for $T_{\mu\nu}=-\Lambda g_{\mu\nu}$, $\Lambda$ being the cosmological constant, we have 
\beq
T_{\mu\nu}U^\nu=-\Lambda U_\mu\,,
\eeq
which vanishes upon projecting orthogonally to $U$.
In this case, a more elaborate force term should be used.
Let us further note that at $w=-1$ there is a sign change in the force, yielding an outward 
push as $w<-1$. This sign change holds the key to the transition from capture to escape scenarios.

\begin{figure}[h]
\begin{center}
\includegraphics[scale=0.3]{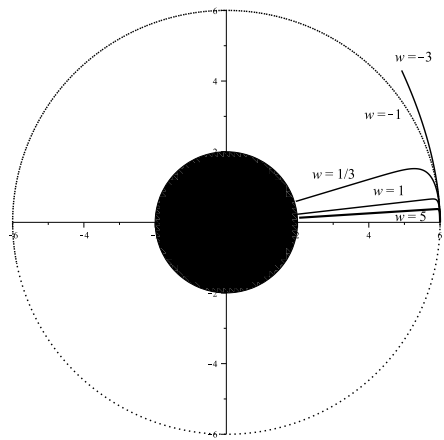}
\end{center}
\caption{Motion of a particle undergoing non-geodesic motion due to pressure forces in 
a Schwarzschild spacetime with a superposed fluid.
The black disk denotes the Schwarzschild horizon, located at $r=2M$. 
The solid lines starting at $r/M=6$ (with $\phi_0=0$, $\nu_0=\nu_K$, $\alpha_0=0$) represent the 
particle motion associated with the scattering of the fluid, with parameters $A=0.1$ and $w=[-3,1/3,1,5]$.  
The short-dashed curve corresponds to $w=-1$, hence to geodesic motion.
With this choice of $A$, all particles scattered by a fluid with $w+1>0$ fall into the hole.
Negative values of $w+1$, instead, allow particles to escape capture, and imply a sudden increase of the speed, which reaches the speed of light, i.e., $\nu=1$, in a finite time
(in order to preserve the causality condition 
for the particle world line, the numerical integration has been stopped at $\nu=1$).
}
\label{fig.1}
\end{figure}

\begin{figure}[h]
\begin{center}
\includegraphics[scale=0.3]{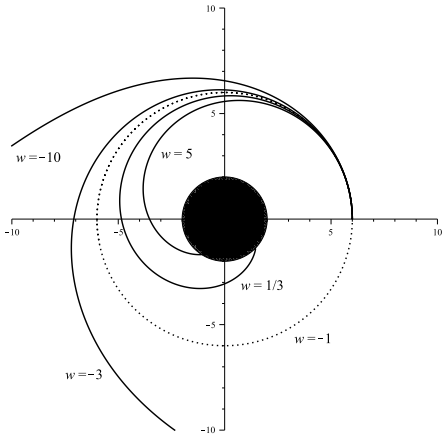}
\end{center}
\caption{The same as in Fig. \ref{fig.1}, but with $A=10^{-3}$ and $w=[-10,-3,1/3,5]$.
With this choice of $A$ (denoting a weaker interaction with respect to the previous case), escaping 
particles ($w<-1$) never reach the speed of light.
}
\label{fig.2}
\end{figure}

Moving to a polar representation of the velocity, i.e., 
$\nu^{\hat r}=\nu\sin{\alpha}$ and $\nu^{\hat \phi}=\nu\cos{\alpha}$, the 
above equations take the following form
\begin{eqnarray}
\label{poleqs}
\frac{d\nu }{d\tau}&=& -A(w+1)\nu   N^{-\frac{(w+1)}{w}}-\gamma^{-1}\sin \alpha \frac{M}{r^2N }\,, \nonumber\\
\frac{d\alpha}{d\tau}&=&\gamma  \frac{N}{r\nu }\cos \alpha (\nu^2-\nu_K^2)\,, \nonumber\\
\frac{d r}{d\tau}&=& \gamma N \nu\sin{\alpha}\,,\qquad
\frac{d \phi}{d \tau}=\frac{\gamma}{r} \nu\cos{\alpha}\,,
\end{eqnarray}
where the interaction parameter
\beq
A=\frac{\sigma}{m}  \rho_0^\infty\,,
\eeq
has been introduced so that $A=0$ identifies the equatorial geodesic case.
Note that the presence of $A$ affects only the first equation of (\ref{poleqs}). 
Apart from very special situations, the analysis of these equations can only be performed numerically.
Figs. \ref{fig.1} and \ref{fig.2} show typical orbits for the same choice of initial conditions, but different values 
of the interaction strength parameter $A$, at different values of the parameter $w$. 
Simple inspection shows that values of $w>-1$ ($w<-1$) imply capture (escape) from the hole, as already stated. 
In Fig. \ref{fig.1}, escaping orbits are characterized by a monotonically increasing speed, up to the speed of light.   
This does not occur in Fig. \ref{fig.2}, due to the very small value of the coupling parameter $A$.
In both cases, however, the deviations from geodesic motion are clearly appreciated.
% --------------------------------------------------------------------------------------
% ------------------------------------------------------------------------------------------------

\subsection{Scattering by multi-component matter}

In principle, it is interesting to consider multi-component fluid scenarios, in which the matter around the 
hole corresponds to regions of different fluids (with different $w$), superposed to the Schwarzschild background. 
This is only an approximation, since the various (ideal) fluids in the different regions 
would rapidly mix and yield a mixture with intermediate values of $w$.
However, if only as a hypothetical speculation, it is of interest to explore the way the equation for $\nu$
(and only that) slightly modifies to account for the presence of multi-component matter, say with 
${\mathcal N}$ components, associated with the parameters
$A_1 \dots A_{\mathcal N}$ and  $w_1\ldots w_{\mathcal N}$.
The new evolution equation for $\nu$ is thus given by  
\begin{eqnarray}
\label{poleqs2}
\frac{d\nu }{d\tau}&=& -\sum_{i=1}^{\mathcal N}  A_i (w_i+1)\nu   N^{-\frac{(w_i+1)}{w_i}}-\gamma^{-1}\sin \alpha \frac{M}{r^2N } \,.
\end{eqnarray}
Explicitly, for ${\mathcal N}=2$ we have, for example,
\begin{eqnarray}
\label{poleqs3}
\frac{d\nu }{d\tau}&=& -A_1 (w_1+1)\nu   N^{-\frac{(w_1+1)}{w_1}}-A_2 (w_2+1)\nu   N^{-\frac{(w_2+1)}{w_2}}\nonumber\\
&& -\gamma^{-1}\sin \alpha \frac{M}{r^2N } \,.
\end{eqnarray}
This is equivalent to a heterogeneous mixture with a radially-dependent $w$ parameter, namely
\beq
w(r)=\frac{A_1 N_1(r) w_1 + A_2 N_2(r) w_2}{A_1 N_1(r) + A_2 N_2(r)}\,,
\eeq
where we have set $N_k(r) \equiv N(r)^{-(1+w_k)/w_k}$.
One may then consider the presence of both ordinary or exotic matter, with special choices of $w_1$ and $w_2$. 
Numerical integration of the orbits in this case, confirms the previous analysis, namely that
the dominating value of $w$ determines the final fate, capture or escape, of the particle trajectory.
A more in depth analysis of the multi-component fluid scenario will make the object of future investigations.

\section{Concluding Remarks}

The motion of massive test particles interacting with a perfect fluid superposed to a Schwarzschild black hole is investigated.
The equation of state describing the fluid is taken as $p_0=w\rho_0$, which corresponds to different physical scenarios for different walues of $w$, from ordinary matter to exotic matter. 
The interaction with the fluid distribution is modeled by a force term entering the equations of motion given by the 4-momentum density observed in the particle's rest frame with a multiplicative constant factor expressing the strength of the interaction itself.
Deviations from geodesic motion are analyzed for different values of the scattering parameter as well as of the parameter $w$.
The latter turns out to be crucial in distinguishing among capture ($w+1>0$) or 
escape ($w+1<0$) from the black hole.
The measurement of such effects may offer a new means of gleaning information on
the nature of the cosmic matter surrounding astrophysical objects.

\end{document}